\documentclass[aps,prc,superscriptaddress,twoside,twocolumn,nofootinbib,showpacs, 10pt]{revtex4-1}
\usepackage{float}
\usepackage{amsmath,amssymb}
\usepackage{mathtools}
\usepackage{bbold}
\usepackage{physics}
\usepackage[usenames, dvipsnames]{xcolor}
\usepackage{braket}
\usepackage{graphicx,bm}
\usepackage{slashed}
\usepackage{booktabs}
\usepackage{tensor}
\usepackage{multirow}
\usepackage{mwe}
\usepackage{changes}
\usepackage{enumerate}
\usepackage{bigints}
\usepackage{subfigure}
\usepackage{float}

\usepackage{tikz}
\usetikzlibrary{arrows.meta}
\usetikzlibrary{shapes.multipart}
\tikzset{%
  >={Latex[width=2mm,length=2mm]},
            base/.style = {rectangle, rounded corners, draw=black,
                           minimum width=2.5cm, minimum height=1.2cm,
                           text centered},
  activityStarts/.style = {base, fill=blue!30},
       startstop/.style = {base, fill=red!30},
    activityRuns/.style = {base, fill=green!30},
         process/.style = {base, minimum width=2.5cm, fill=orange!15},
}

\allowdisplaybreaks

\newcommand{\pvec}[1]{\vec{#1}\mkern2mu\vphantom{#1}}

\allowdisplaybreaks

\begin{document}

\title{Application of normalizing flows to nuclear many-body perturbation theory}

\author{Pengsheng Wen}
\email{pswen2019@physics.tamu.edu}
\affiliation{Cyclotron Institute, Texas A\&M University, College Station, TX 77843, USA}
\affiliation{Department of Physics and Astronomy, Texas A\&M University, College Station, TX 77843, USA}

\author{Jeremy W. \surname{Holt} }
\email{holt@physics.tamu.edu}
\affiliation{Cyclotron Institute, Texas A\&M University, College Station, TX 77843, USA}
\affiliation{Department of Physics and Astronomy, Texas A\&M University, College Station, TX 77843, USA}

\author{Albany Blackburn}
\email{ablackburn@g.hmc.edu}
\affiliation{Harvey Mudd College, Claremont, CA 91711, USA}

\date{\today}

\begin{abstract}
Many-body perturbation theory provides a powerful framework to study the ground state and thermodynamic properties of nuclear matter as well as associated single-particle potentials and response functions within a systematic order-by-order expansion. However, computational challenges can emerge beyond the lowest orders of perturbation theory, especially when computing both single-particle potentials and response functions, which in general are complex-valued and require Cauchy principal value calculations of high-dimensional integrals. We demonstrate that normalizing flows are suitable for Monte Carlo importance sampling of both regular and irregular functions appearing in nuclear many-body calculations. Normalizing flows are a class of machine learning models that can be used to build and sample from complicated distributions through a bijective mapping from a simple base distribution. Furthermore, a well-trained model for a certain target integrand can be efficiently transferred to calculate related integrals with varying physical conditions. These features can enable more efficient tabulations of nuclear physics inputs to numerical simulations of supernovae and neutron star mergers across varying physical conditions and nuclear force models.
\end{abstract}

\maketitle

\section{Introduction}
Theoretical modeling of the nuclear matter equation of state (EOS) at both zero and finite temperature is important for understanding neutron star structure, the cooling of neutron stars, and the dynamical evolution of core-collapse supernovae and neutron star mergers \cite{shen91,lattimer91,yakovlev_neutron_2005,constantinou17,brown_rapid_2018,sumiyoshi_equation_2021,annala_multimessenger_2022,kumar23,du19}. Besides the EOS, nuclear matter response functions and single-particle potentials also provide important information about the system. In particular, response functions characterize how the medium couples to external probes, which is important for understanding neutrino opacities in supernovae and neutron star mergers \cite{iwamoto_effects_1982,sawyer89,burrows98,reddy98,kotake_impact_2018,lovato_neutral-current_2014,pastore_linear_2014,shen_neutrino_2003,shin24} as well as the energy density in neutron star inner crusts, where a gas of unbound neutrons interacts with a lattice of neutron-rich nuclei \cite{horowitz05}. Nucleon single-particle potentials and effective masses \cite{li18}, which can be obtained from the nucleon self-energy, directly impact charged-current neutrino reactions in supernovae and neutron star mergers \cite{roberts12,martinez-pinedo12,rrapaj15} as well as nuclear matter thermodynamics and the evolution of core-collapse supernovae \cite{yasin_equation_2020,donati_temperature_1994}.

The accurate and consistent calculation of various nuclear matter properties (the EOS, self-energy, and response functions) poses a significant theoretical challenge but is especially important in dynamical simulations of supernovae and neutron star mergers, where inconsistencies in the treatment of the EOS and neutrino opacities can lead to simulation artifacts. Phenomenological nuclear mean field models, which employ effective interactions fitted to the properties of nuclei and nuclear matter, have long been the method of choice for tabulating the equation of state and neutrino opacities for astrophysical simulations due to their computational simplicity and their treatment of nuclear matter and finite nuclei on an equal footing \cite{sotani_gravitational_2022,ishizuka_tables_2008,oertel_hyperons_2015,steiner_core-collapse_2013,hempel_statistical_2010}. However, beyond-mean-field effects contribute meaningfully to the nuclear single-particle potential and the associated nucleon effective mass \cite{tan_mean-field_2016,zuo_interplay_2002,zuo_microscopic_2002}. In addition, higher-order corrections to the density-dependent isospin asymmetry energy \cite{wen_constraining_2021,wellenhofer_divergence_2016} only appear beyond the mean field level. Therefore, microscopic approaches starting from fundamental nucleon-nucleon and many-nucleon forces may be needed for a realistic description of many-body effects crucial for astrophysical applications.

Nonperturbative many-body methods, such as quantum Monte Carlo imaginary-time evolution, have had success in extracting the equation of state and static response functions \cite{akmal98,gezerlis13,wlazlowski14,tews16,buraczynski16,alexandru21,ma24}. However, such calculations are very computationally expensive, which generally prohibits a full exploration of systematic uncertainties associated with e.g., the choice of nuclear potential. In addition, it may be difficult to extract dynamical quantities such as the nucleon self-energy \cite{holtzmann23} and the energy dependence of response functions \cite{roggero13}. For these reasons, many-body perturbation theory provides an appealing framework for computing on a consistent footing the nuclear matter equation of state at zero- and finite-temperature, single-particle potentials, and dynamical response functions, all in the thermodynamic limit and for arbitrary proton-to-neutron ratio.

In many-body perturbation theory calculations of nuclear matter properties, one encounters multidimensional integrals of increasing complexity. While the nuclear equation of state is a real-valued function of the density $n$, temperature $T$, and proton fraction $Y_p$, the nucleon self-energy $\Sigma(q,E)$ is complex valued and depends on the momentum $q$ and energy $E$ of the propagating particle (in addition to the local thermodynamic variables $n, T, Y_p$). Similarly, nuclear matter response functions $\chi(q,\omega)$ are complex-valued and depend on both the energy $\omega$ and momentum $q$ transferred to the medium \cite{iwamoto_effects_1982,davesne_effect_2015,shen_spin_2013}.

In previous work \cite{brady_normalizing_2021}, we have shown that normalizing flow machine learning models \cite{tabak10,tabak13,papamakarios19} can yield dramatic improvements in Monte Carlo importance sampling calculations of the finite-temperature nuclear equation of state. Compared to state-of-the-art importance sampling algorithms \cite{Hahn_2005,Lepage2020AdaptiveMI} that have previously been used in many-body perturbation theory calculations of the equation of state \cite{drischler_chiral_2019}, normalizing flows exhibited an improvement in the sampling efficiency by at least a factor of 100. The precision of importance sampling integration is strongly dependent on how well the sampling distribution matches the normalized absolute value of the integrand. Normalizing flows map a simple base distribution (e.g., a uniform or Gaussian distribution) to a more complicated target distribution through a series of change-of-variable transformations learned via neural networks. Once the flow model is constructed, it is simple to pass samples generated from the base distribution to the target distribution with the appropriate statistical weight. Because of their ability to build highly expressive models of target distributions, normalizing flows have been a widely used tool for density estimation \cite{tabak_family_2013}, variational inference and generative modeling of images \cite{kingma_glow_2018,yang_pointflow_2019}. Normalizing flows have also attracted attention outside of the computational science community \cite{wirnsberger_targeted_2020,kanwar_equivariant_2020,gao_event_2020,hajij_normalizing_2022,wen23} as sample generators.

In the present work, we explore normalizing flows as a comprehensive tool for many-body perturbation theory calculations of nuclear matter properties. In particular, we focus on whether the method is suitable for calculations of nucleon self-energies and nuclear matter response functions, which are complex-valued and in the most general case depend on five independent variables ($n, T, Y_p, E, q$). Since the computational challenges are essentially the same for these two nuclear matter quantities, we will focus on response functions for which even the zeroth-order many-body perturbation theory contribution is complex-valued and requires a Cauchy principal value integration to extract the real part. In previous work, we have shown that a well-trained normalizing flow model can be efficiently transferred to nearby phase space points in ($n,T,Y_p$). In the present work, we also show that well-trained models can be efficiently transferred to different kinematic regimes in ($E,q$) space, relevant for single-particle potentials and response functions. Finally, we investigate the efficiency of different loss functions in the machine learning algorithm as well as quasi-random number generators that more uniformly cover the domain space.

The paper is organized as follows. In Section \ref{sec:importance_sampling} we give a brief introduction to Monte Carlo integration based on importance sampling via normalizing flows. In Section \ref{results} we investigate the application of normalizing flows to the calculation of the second-order perturbation theory contribution to the nuclear matter grand canonical potential. Particular attention is paid to the efficiency of training the normalizing flow for different choices of random number generators and loss functions. We also study in this section the zeroth-order perturbative contribution to the density-density response function of nuclear matter and the ability of normalizing flows to accurately model integrands with poles in the integration region. The paper ends with a summary and conclusions.

\section{Monte Carlo Importance Sampling with Normalizing Flows}\label{sec:importance_sampling}
Monte Carlo importance sampling is widely used for the calculation of high-dimensional integrals. The key idea is that 
the integral of a function $\psi(\pvec{x})$ over the domain $\mathcal{D}$ can be written as
\begin{equation}
	I 
	= \int_{\mathcal{D}} \psi(\pvec{x})\, d\! \pvec{x}
	= \int_{\mathcal{D}} \frac{\psi(\pvec{x})}{p(\pvec{x})} 
	p(\pvec{x})\,  d\! \pvec{x},
\end{equation}
where we have introduced an arbitrary probability density function (PDF) $p(\vec{x})$. This equation can be interpreted as the expectation value of the function $\frac{\psi(\vec{x})}{p(\vec{x})}$ of the random variable $\vec{x}$ with probability density $p(\pvec{x})$. Then an integral estimate is obtained by drawing samples from the PDF $p(\vec{x})$ and evaluating
\begin{equation}
I \approx \langle I \rangle = 
\frac{1}{N}
\sum_{i=1}^N \frac{\psi\qty(\pvec{x}_i)}{p\qty(\pvec{x}_i)}, 
\quad x_i \in p(\pvec{x})
\label{pav}
\end{equation}
with standard deviation
\begin{align}
	\sigma^2 
	&=
	\qty(
	\mathbf{E}_p \qty[
	\qty(\frac{\psi(\pvec{x})}{p(\pvec{x})})^2 ]
	- 
	\qty(
	\mathbf{E}_p \qty[
	\frac{\psi(\pvec{x})}{p(\pvec{x})}])^2
	)\!\Bigg/\!(N-1).
 \\
 &\approx 
 \left(\frac{1}{N-1}\right)^2 \sum_{i=1}^{N}
 \left(\frac{\psi(\pvec{x}_i)}{p(\pvec{x}_i)}
 - \frac{1}{N}\sum_{j=1}^{N}\frac{\psi(\pvec{x}_j)}{p(\pvec{x}_j)}\right)^2
 \label{eq:sig}
\end{align}
According to Eq.\ \eqref{eq:sig}, there are two approaches to improve the precision of the calculation. The first is to increase the number of samples, since $\sigma$ is approximately proportional to $\sqrt{1/N}$. This approach, however, is often inefficient since an order of magnitude improvement in the integral estimate requires an increase in the batch size by a factor of 100. The second approach is to draw samples from a PDF that is as close as possible to the normalized absolute value of the integrand $|\psi(\pvec{x})|$. When $|\psi(\pvec{x})|/p(\pvec{x})$ is a constant everywhere, the standard deviation in Eq.\ \eqref{eq:sig} vanishes. In this case, more samples are generated at locations where the magnitude of the integrand is largest. The challenge is how to efficiently generate samples according to the distribution $|\psi(\pvec{x})|/\tilde{I}$ or a different PDF that is very close to $|\psi(\pvec{x})|/\tilde{I}$, where $\tilde{I}$ is the integral normalizing constant.

In recent years, normalizing flows have developed into a useful machine learning algorithm to perform Monte Carlo importance sampling integration. The key steps to constructing a normalizing flow for Monte Carlo sampling of multidimensional integrals are as follows: (a) start by defining the base PDF, which should be sufficiently simple that samples can efficiently be drawn from it; (b) define the transformation function used to map the base PDF to a more complicated one; (c) construct a sequence of ``maskings'', which define how the transformation function applied to a given coordinate depends on the other coordinates; and (d) train the full transformation such that the transformed and target distribution are as close as possible.

\paragraph{Transformation of a PDF:}
In our work, we start with a set of samples $\{\pvec{x}^{(0)}_i\}$ drawn from a uniform distribution $q_0$ within the integration region:
\begin{align}
	q_0(\pvec{x}^{(0)}) = \frac{1}{V},
\end{align}
where $V$ represents the multidimensional volume of the integration domain. Given a change of variables transformation function $g_0$ defined by
\begin{align}
	\pvec{x}^{(1)}_i = g_0(\pvec{x}_i^{(0)}),
\end{align}
the transformed PDF is obtained from
\begin{align}
	I 
	&= 
	\int_{\mathcal{D}} 
	f\qty(\pvec{x}^{(0)}) 
	q_0\qty(\pvec{x}^{(0)}) d \pvec{x}^{(0)}
	\nonumber\\
    &= 
	\int_{\mathcal{D}} 
	f\qty(g_0^{-1}(\pvec{x}^{(1)}))
	q_0\qty(g_0^{-1}(\pvec{x}^{(1)}))
	\qty|\frac{\partial g_0^{-1}}{\partial \pvec{x}^{(1)}}| 
	d \pvec{x}^{(1)},
	\nonumber\\
\end{align}
where $|\partial g_0^{-1}/ \partial \pvec{x}^{(1)}|$ is the absolute value of the determinant of the Jacobian matrix. The above equation indicates that the PDF of the transformed samples is given by
\begin{equation}
\label{pdf2}
	q_1 (\pvec{x}^{(1)}) = 
	q_0(\pvec{x}^{(0)})
	\qty| \frac{\partial g_0 }{\partial \pvec{x}^{(0)}} |^{-1}.
\end{equation}
To build a new PDF that is close to the target distribution $|\psi(\pvec{x})| / \tilde{I}$, a single transformation function is often insufficient, and therefore several transformation layers are necessary:
\begin{align}
	\pvec{x}^{(L)}_{i}
	&= 
	g_{L} \qty( g_{L-1}\qty( \dots g_{0}\qty(\pvec{x}_i^{(0)}))),
	\label{eq:coupling}
\end{align}
where $L+1$ is the total number of layers. The final PDF is then given by
\begin{align}
	q_{L}\qty(\pvec{x}^{(L)})
	&= 
	q_{0}\qty(\pvec{x}^{(0)})
	\prod_{i=0}^{L} 
	\qty| \frac{\partial g_{i}}{\partial \pvec{x}^{(i)}} |^{-1}.
	\label{eq:tfPDF}
\end{align}

\paragraph{Coupling transformations:}
From the above discussion, we see that in order to develop a normalizing flow that can efficiently be learned via neural networks, the transformation functions $g_i$ should be invertible and have computationally inexpensive Jacobian matrices. In the present work, we employ coupling transforms for this purpose. The basic idea of the coupling transform is that at each transformation layer, some dimensions of the sample are chosen to be the input variables of the neural network (NN) that generates the parameters for the transformation functions for the remaining dimensions. For instance, one such coupling layer might have the form:
\begin{align}
	\pvec{x}
	&=
	(x_1, x_2, \dots, x_d)
	= \pvec{x}_I + \pvec{x}_T,
	\nonumber\\
	\pvec{x}_I
	&= (x_1, \dots, 0, \dots, x_j, \dots x_d),
	\nonumber\\
	\pvec{x}_T
	&= (0,	\dots, x_i, \dots, 0, \dots  0 ).
\end{align}
The partitioning of the sample into a set of invariant dimensions $\pvec{x}_I$, which remain unchanged at the present transformation layer, and the transformed dimensions $\pvec{x}_T$ is called a ``masking''. By combining multiple transformation layers with different masking schemes, it is possible to ensure that every dimension can influence the flow of the other dimensions. This allows for the construction of highly expressive probability densities with correlations among the different dimensions.

For a given masking, we have two arrays of numbers, $\mathbf{x}_I$ and $\mathbf{x}_T$, which store all of the nonzero elements of $\pvec{x}_I$ and $\pvec{x}_T$ separately as
\begin{align}
	\mathbf{x}_I
	& \equiv [ x_1, \dots, x_i, \dots, x_d ]
	\\
	\mathbf{x}_T
	& \equiv [ \dots, x_j, \dots ].
\end{align}
In addition, we introduce two arrays that store the indices of the dimensions in $\mathbf{x}_I$ and $\mathbf{x}_T$ as
\begin{align}
	D_I = [1, \dots, i, \dots, d], \quad 
	D_T = [\dots, j, \dots].
\end{align}
Neural networks are then applied to generate e.g., the parameters $\vec{\theta}_j$ of the transformation function $r_j$ acting on the coordinate $x_j \in \mathbf{x}_T$, with $\mathbf{x}_I$ the input of the neural network transformation $\vec {\rm N}_j$:
\begin{align}
	\{
		\vec{\theta}_j| j \in D_T
	\}
	&= 
	{\vec {\rm N}}_j (\mathbf{x}_I).
\end{align}
Then 
\begin{align}
	x_j \to r_j({\vec {\rm N}}_j(\mathbf x_I); x_j).
\end{align}
Notice that this transformation does not change some dimensions of a sample. Since none of the transformed dimensions couple to each other at a given layer, the Jacobian matrix used to calculate the transformed PDF can be arranged into a triangular structure
\begin{align}
	\frac{\partial g_i}{\partial x_j}
	&= 
	\begin{cases}
		\begin{aligned}
		&\delta_{i, j}, 
		\quad  &i \in D_I \\
		&\frac{\partial \, r_i}{\partial x_j},
		\quad & j \in D_T \quad {\rm and} \quad i \in D_T \\
  		&\sum_k \frac{\partial \, r_i}{\partial {\rm N}_{i,k}}\frac{\partial {\rm N}_{i,k}}{\partial x_j},
		\quad & j \in D_I \quad {\rm and} \quad i \in D_T.
		\end{aligned}
	\end{cases}
	\label{eq:JacMatrix}
\end{align}


\paragraph{Rational Quadratic Splines:}
There are several convenient choices for the transformation functions $r_j$. In the present work, we employ rational quadratic splines \cite{durkan_neural_2019,11.1093/imanum/2.2.123}. For simplicity, we start with an illustrative example in one dimension. Suppose the initial region of $x$ is $(x_l, x_u)$ and the final region after mapping is $(y_l, y_u)$. We can use $K+1$ monotonically increasing knots $(x_k, y_k)$, where $(x_1, y_1) = (x_l, y_l)$ and $(x_{K+1}, y_{K+1}) = (x_u, y_u)$, to split the regions into $K$ bins. The width and height of the $k$-th bin are $w_k = x_{k+1} - x_k$ and $h_k = y_{k+1} - y_{k}$. We denote the average slope over the $k$-th bin as $s_k \equiv h_k / \omega_k$. In addition, we specify the derivatives $\delta_k$ at each knot. Then the rational-quadratic transformation of a sample $x$ is given by
\begin{align}
	y 
	&= r(x)
	= 
	\sum_{k=1}^{K}
	\frac{\alpha_k(x)}{\beta_k(x)}
	\left[ \theta (x - x_k ) - \theta (x - x_{k+1}) \right],
	\\
	\alpha_k(x) 
	&=
	s_k y_{k+1} \xi^2 + \left( y_k \delta_{k+1} + y_{k+1} \delta_k\right) 
	\xi (1-\xi),
	\nonumber\\
	&\quad + s_k y_k (1-\xi)^2
	\\
	\beta_k(x)
	&= 
	s_k \xi^2 + \left( \delta_{k+1} + \delta_k \right) \xi(1-\xi) 
	+ s_k (1-\xi)^2,
		\label{eq:tffunc}
\end{align}
where $\xi = ( x - x_k) / w_k $ and $\theta(x)$ is the step function. The rational-quadratic spline transformation therefore depends only on the set of $\vec {\theta}$ parameters consisting of the $K+1$ derivatives $\vec {\theta}^{\, \delta}$ at each knot, the $K$ bin widths $\vec{\theta}^{\, w}$, and the $K$ bin heights $\vec{\theta}^{\, h}$. With these $3K+1$ parameters, we build a single transformation function partitioned into $K$ bins to cover the total integration region of $x$.

Two additional important features of the rational quadratic spline transformation are (i) it is invertible and (ii) it has a derivative that is easily computable. In particular, the derivative of the above transformation function within the $k$-th bin is given by
\begin{align}
	\frac{d}{dx} y 
	&= 
	\frac{s_k^2\qty[ \delta_{k+1} \xi^2 + 2s_k \xi (1 - \xi) 
	+ \delta_k (1-\xi)^2]}
	{\qty[s_k + \qty(\delta_{k+1} + \delta_k - 2s_k) \xi (1-\xi)]^2}.
	\label{eq:dydx}
\end{align}
The rational-quadratic spline transformation can easily be extended to the multi-dimensional case, where every dimension has its own independent transformation function, Eq.\ \eqref{eq:tffunc}, 
which allows us to transform all the dimensions of $\vec{x}$ one by one to $\vec{y}$.

\begin{figure*}[htp]
\centering
\resizebox{!}{0.62\paperheight}{
\begin{tikzpicture}[node distance=2.2cm,
    every node/.style={fill=white}, 
    every text node part/.style={align=center}]
  \node (start)           [activityStarts]
        {Sample Generator};
  \node (initialsample)   [process, below of = start]
        {Initial Samples: 
        $\{\vec{x}^{\,(0)}\} \in q^{(0)}$};
  \node (tfele)           [process, below of = initialsample]
        {$\mathbf{x}^{\,(0)}_T$};
  \node (layer1)          [right of=tfele, xshift = -5.cm]
        {Layer 1};
  \node (identityele)     [process, right of=tfele, xshift = 3.5cm]
        {$\mathbf x^{\,(0)}_I$};
  \node (parameter1)      [activityRuns, below of=identityele]
        {$\vec{\theta}^{\,(0)}_j = {\vec {\rm N}}^{(0)}_j(\mathbf x^{(0)}_I)$};
  \node (tffunc1)         [process, below of=tfele]
        {Transformation function \\ ${x^{(1)}_{T,j}}=r_j^{(0)}\qty(\vec{\theta}^{\,(0)}_j; x^{(0)}_{T,j})$};
  \node (tfsample1)       [process, below of=tffunc1]
        {Transformed samples 1 \\ 
        $\{ \vec{x}^{\,(1)}
         = \mathbf{x}^{\,(1)}_{T} \bigoplus \mathbf{x}^{\,(0)}_{I}\}
        \in q^{(1)}$};
  \node (flayer)          [process, below of=tfsample1]
        {Multiple layers};
  \node (tfsample)        [process, below of=flayer]
        {Transformed samples $L$ \\ 
        $\{ \vec{x}^{\,(L)}
         = \mathbf{x}^{\,(L)}_{T} \bigoplus \mathbf{x}^{\,(L-1)}_{I}\}
        \in q^{(L)}$};
  \node (integral)       [process, below of=tfsample]
        {Estimation of the integral};
  \node (loss)       [activityRuns, right of=integral, xshift=2.5cm]
        {Loss function};
  \node (adam)       [activityRuns, right of=loss, xshift=2.5cm]
        {Adam Optimizer};
  \draw[->] (start) -- (initialsample);
  \draw[->] (initialsample) -- node{mask1}(tfele);
  \draw[->] (initialsample) -| node{mask1}(identityele);
  \draw[->] (tfele) -- (tffunc1);
  \draw[->] (identityele) -- (parameter1);
  \draw[->] (parameter1) -- (tffunc1);
  \draw[->] (identityele) -- (tfsample1);
  \draw[->] (tffunc1) -- (tfsample1);
  \draw[->, dashed] (tfsample1) -- (flayer);
  \draw[->, dashed] (flayer) -- (tfsample);
  \draw[->] (tfsample) -- (integral);
  \draw[->] (integral) -- (loss);
  \draw[->] (loss) -- (adam);
  \draw[->] (adam) -- (flayer);
  \draw[->] (adam) -- (parameter1);
\end{tikzpicture}
}
\caption{Work flow for normalizing flows. A sample $\vec{x}^{(0)}$ is generated from a uniform distribution and moves through the flow. At the end of the flow, the integral is estimated and the loss function is calculated. The loss function is then used to optimize the neural networks via the Adam gradient descent optimizer.}
\label{fig:nflowworkflow}
\end{figure*}
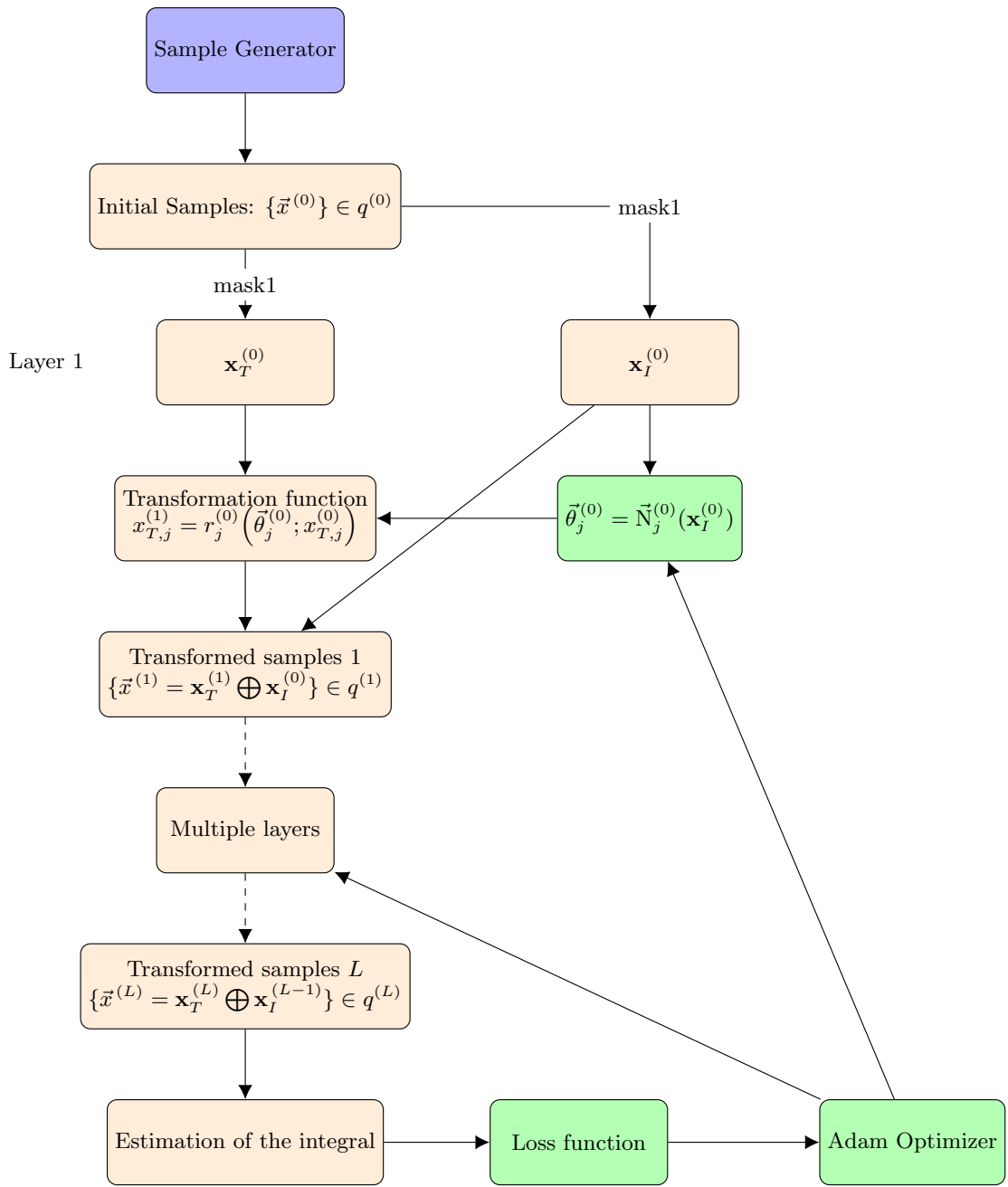

\paragraph{Loss Function:}
The training of the neural networks that determine the rational quadratic spline transformation parameters is carried out by comparing the final transformed PDF $p(\pvec{x})$ to the target PDF $|\psi(\pvec{x})|/\tilde{I}$. We introduce a loss function to evaluate this difference. Due to the fact that there is no unique way to define how close one PDF is to another, there are several kinds of loss functions that can be employed for this purpose (see e.g., the Appendix of Ref.\ \cite{gao_i-_2020}). For example, the Pearson $\chi^2$ divergence is defined as 
\begin{align}
	D_{\chi^2}
	&= 
	\bigintss \frac{\qty(|\psi(\pvec{x})|/\tilde{I} - p(\pvec{x}))^2}
	{p(\pvec{x})} d \pvec{x}
	\nonumber\\
	&\approx
	\frac{1}{N}
	\sum_{i=1}^{N}
	\frac{\qty(|\psi(\pvec{x_i})|/\tilde{I} - p(\pvec{x}_i ))^2}
	{p^2(\pvec{x}_i)}, 
	\quad \pvec{x}_i \in p(\pvec{x}),
\end{align}
where $\tilde{I}$ is the value of the integral with integrand $|\psi(\pvec{x})|$, which can be estimated by 
\begin{equation}
\label{mcie}
\tilde I \approx 
\frac{1}{N}
\sum_{i=1}^N \frac{\left | \psi\qty(\pvec{x}_i) \right |}{p\qty(\pvec{x}_i)}, 
\quad \vec x_i \in p(\pvec{x}).
\end{equation}
Because the neural networks generate all of the parameters to build the flows, the loss function is dependent on the NNs. At each iteration, the NNs are called to generate parameters to build the flows. A batch of samples is generated from the base distribution and transformed by the flows. The transformed samples and PDF are then used to estimate the integral and the loss function. After the loss function is calculated, the NNs get updated using the gradient descent optimization algorithm Adam \cite{kingma2017adam}. The construction of the NNs and optimization are completed by the package PyTorch \cite{NEURIPS2019_9015} in this work.
The workflow is illustrated in Figure \ref{fig:nflowworkflow}

Normalizing flows have two important advantages for Monte Carlo importance sampling of high-dimensional integrals. The first is expressibility. The parameters in the transformation functions generated by the NNs can be efficiently optimized through training. After sufficient iterations, the transformed PDF can be very close to the target distribution, which enables one to estimate the integral with high precision. As an example, in previous work \cite{brady_normalizing_2021} we showed that 5000 samples drawn from a uniform distribution over a 7-dimensional domain space and transformed via a normalizing flow could result in a relative precision of $\sigma / I \simeq 2\times 10^{-3}$ for a second-order perturbation theory calculation of the finite-temperature equation of state. The second advantage is transferability. For example, in previous work \cite{brady_normalizing_2021} we showed that a NN model trained to calculate the equation of state at one density-temperature phase-space point can then be transferred to a nearby phase space point with little loss of precision. In addition, changing the model of the nuclear force also did not result in a significant loss of precision. Transferability is particularly important for the efficient calculation of nuclear matter properties that depend on multiple thermodynamic variables (e.g., $n,T,Y_p$) and dynamical variables (e.g., energy transfer $\omega$ and momentum transfer $q$ for response functions). 

\section{Results}
\label{results}
In this section, we apply normalizing flows to the calculation of the second-order perturbation theory contribution to the nuclear matter grand canonical potential as well as the zeroth-order density-density response function of pure neutron matter. First, we explore the efficiency of normalizing flow importance sampling when varying the choice of sample generator and loss function. To investigate the sample generator efficiency, we employ a pseudo-random number generator as well as three different quasi-random number generators, which are designed to more evenly cover the domain space. In addition, we study eight different loss functions for training the normalizing flow neural network. We then turn our attention to the applicability of normalizing flows to treat the complex-valued density response function, whose real part in general requires the evaluation of principal value integrals.

\subsection{Grand Canonical Potential and the EOS}
\label{sec:GrandCanonicalPot}

The grand canonical potential $\Omega$ is a fundamental thermal statistical quantity. 
It is defined according to
\begin{align}
	\Omega(T, V, \mu)
	&= 
	E - TS - \mu N,
\end{align}
where $E$ is the internal energy, $T$ is the temperature, $S$ is the entropy, $\mu$ is the chemical potential, $V$ is the volume, and $N$ is the number of particles of the system. From statistical mechanics, the grand canonical potential can also be calculated through the partition function $Z_G$:
\begin{align}
	Z_G
	&= \Tr e^{-\beta (\hat{H} - \mu \hat{N})}
	= e^{-\beta \Omega}, 
\end{align}
where $\beta = 1/T$ is the inverse of temperature, $\hat{H}$ is the Hamiltonian, and $\hat{N}$ is the particle number operator. The trace operator denotes summing over all possible quantum states of the system. The order-by-order calculation of the grand canonical potential from perturbation theory is given by \cite{kohn_ground-state_1960,luttinger_ground-state_1960}:
\begin{align}
	\Omega
	&= 
	\Omega^{(0)}
	+ 
	\sum_{i\ge 1} \Omega^{(i)},
	\nonumber\\
	\Omega^{(i)}
	&= 
	\frac{1}{\beta}
	\sum_{n\ge 1}
	\frac{(-1)^{n+1}}{n!}
	\int_0^\beta d\tau_1
	\int_0^\beta d\tau_2
	\dots 
	\int_0^\beta d\tau_n
	\nonumber\\
	&\quad \times
	\langle T_\tau 
	[\hat K_I(\tau_1) \hat K_I(\tau_2) \dots \hat K_I(\tau_n)]
	\rangle_{L},
	\label{eq:omega}
\end{align}
where $\Omega^{(i)}$ is the $i$th-order contribution to the grand canonical potential and $\hat{K}_I(\tau)$ is the $\tau$-dependent interaction part of the grand canonical Hamiltonian. The subscript $L$ on the bracket denotes that only linked diagrams contribute in the expansion.


\begin{figure}
	\centering
	\centering
	\includegraphics[width=0.5\linewidth]{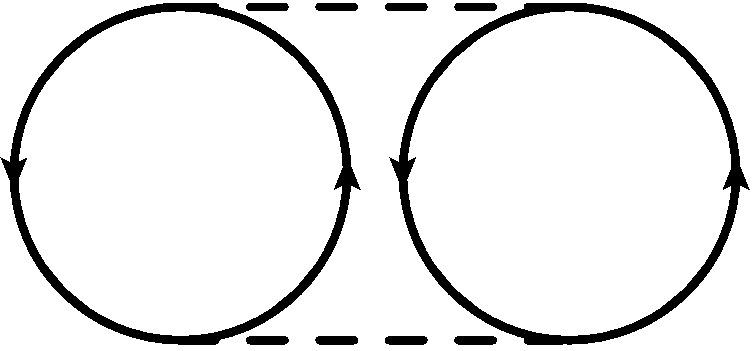}
	\caption{The diagrammatic 2nd-order normal contribution to the grand canonical potential, $\Omega^{(2)}$.}
	\label{fig:2ndomega}
\end{figure}

In this work, we consider the second-order normal contribution to the grand canonical potential $\Omega^{(2)}$ in Eq.~\eqref{eq:omega} \cite{wellenhofer_isospin-asymmetry_2017,wellenhofer_nuclear_2014}, which has the form
\begin{align}
\label{omega2}
	\Omega^{(2)}
	&= \!
	-\frac{1}{8} \!
	\sum_{1,2,3,4} \!
	\langle 12 | V | 34 \rangle 
	\langle 34 | V | 12 \rangle 
	\frac{f_1 f_2 \bar{f}_3 \bar{f}_4 - \bar{f}_1 \bar{f}_2 f_3 f_4}
	{\epsilon_3 + \epsilon_4 - \epsilon_1 - \epsilon_2},
\end{align}
where $\epsilon_i = \pvec{k}_i^2/2M$ is the energy of the $i$-th particle and $f_i$ is the Fermi-Dirac distribution function
\begin{align}
	f_i = \frac{1}{e^{\beta(\epsilon_i - \mu)} + 1}
\label{fdir}
\end{align}
and $\bar{f}_i = 1 - f_i$. 

We first consider a toy model of the nuclear potential from scalar-isoscalar boson exchange given by
\begin{align}
	V({q})
	&= 
	\frac{g^2}{m_\phi^2 + {q}^2},
\label{obe}
\end{align}
where ${q}$ is the magnitude of the momentum exchange between interacted particles, $m_\phi$ is the mass of the exchanged boson, and $g$ is the coupling constant of the interaction. We have shown in previous work \cite{brady_normalizing_2021} that a more complicated model of the nucleon-nucleon potential based on one-pion-exchange and multi-pion-exchange processes does not impact the effectiveness of the normalizing flow importance sampling. To regularize the divergent loop integral in Eq.\ \eqref{omega2}, we introduce
a cutoff function $\cal{R}$ on the momentum-space matrix elements of the potential \cite{10.3389/fphy.2020.00100}:
\begin{align}
	& \langle 12 | V | 34 \rangle
	\to 
	\langle 12 | V | 34 \rangle 
	\mathcal{R}(p, p^\prime),
	\\
	& \mathcal{R}(p, p^\prime)
	= 
	\exp\Bigg[
	- \qty(\frac{p}{\Lambda})^6
	- \qty(\frac{p^\prime}{\Lambda})^6
	\Bigg],
\end{align}
where $p = \frac{1}{2}|\vec k_1 - \vec k_2|$ and $p^\prime = \frac{1}{2}|\vec k_3 - \vec k_4|$ are the incoming and outgoing relative momenta of the two nucleons, and $\Lambda$ is the cutoff scale.
This interaction leads to the following 2nd-order contribution to the grand canonical potential \cite{brady_normalizing_2021}:
\begin{align}
	\Omega^{(2)}
	&= 
	- \frac{Mg^4}{64 \pi^8}
	\int_0^\infty dk_1
	\int_0^\infty dk_2
	\int_0^\infty dk_3
	\nonumber\\
	&\quad \times
	\int_0^\pi d\theta_1
	\int_0^\pi d\theta_2
	\int_0^{2\pi} d\phi_1 
	\int_0^{2\pi} d\phi_2
	\nonumber\\
	&\quad \times 
	k_1^2 k_2^2 k_3^2 \sin\theta_1 \sin\theta_2
	\frac{f_1 f_2 \bar{f}_3 \bar{f}_4 - \bar{f}_1 \bar{f}_2 f_3 f_4}
	{k_1^2 + k_2^2 - k_3^2 - k_4^2}
	\nonumber\\
	&\quad \times 
	\qty[
	\frac{4}{(m_\phi^2 + q_1^2)^2}
	-
	\frac{1}{(m_\phi^2 + q_1^2)(m_\phi^2 + q_2^2)}
	]
	\nonumber\\
	&\quad \times 
	e^{-(p/\Lambda)^6 - (p'/\Lambda)^6},
	\label{eq:2ndOmega}
\end{align}
where $\pvec{k}_4 = \pvec{k}_1 + \pvec{k}_2 - \pvec{k}_3$ from momentum conservation, $ \pvec{q}_1 = \pvec{k}_1 - \pvec{k}_3$, and $ \pvec{q}_2 = \pvec{k}_1 - \pvec{k}_4$. The chemical potential in the above integrand is determined by fixing the temperature and density of the system and then integrating the momentum distribution function, Eq.\ \eqref{fdir}, over all states to obtain the particle number.

In the present work, we consider symmetric nuclear matter, where the densities of protons and neutrons are identical. Evaluating the grand canonical potential at different $(n, T)$ points leads to modifications in the integrand via the Fermi-Dirac distribution functions $f_i$ in Eq.\ \eqref{eq:2ndOmega}. We first train the normalizing flow Monte Carlo importance sampling algorithm at a specific physical condition specified by the density and temperature. We then use the well-trained model to calculate the grand canonical potential at other neighboring conditions as the temperature and density are varied. We study in particular the efficiency of the method as the sampling algorithm and loss function are modified.

\subsection{EOS: Sampling algorithms and loss functions}
In this subsection, we show the performance of the normalizing flow model applied to the 2nd-order grand canonical potential for different choices of the loss function and random number sampling algorithm. The loss function quantifies how well the transformed distribution simulated by the normalizing flow matches the target distribution. The normalizing flow then optimizes the transformation function by minimizing the loss function through training iterations. All of the employed loss functions approach 0 if the two distributions are identical. However, in general they tend to emphasize different properties of the distributions, e.g., their tails, which could influence the efficiency of learning. We implement eight different loss functions, summarized in Table \ref{losst}.

We first train the normalizing flow on the absolute value of the grand canonical potential integrand for a density $n=0.16\,{\rm fm}^{-3}$ and temperature $T=25\,{\rm MeV}$. Since low temperatures lead to a steep integrands due to the Fermi-Dirac distribution functions, it is convenient to begin the training at a large temperature and then allow the model to adjust as the temperature is lowered. The nuclear interaction used to train the models is the simple scalar-isoscalar boson exchange interaction of Eq.\ \eqref{obe}. At each iteration, the Monte Carlo integrators sample points, estimate the value of the integral and give the standard deviation of the estimation. Every iteration can be viewed as an independent estimation. The accumulated estimation collects all of the results up to the last iteration and gives the total result. The accumulated integral value and its standard deviation are given by
\begin{align}
	I_t
	&= 
	\frac{\sum_i I_i / \sigma_i^2}{\sum_i 1/\sigma_i^2},
	\label{eq:weighted_sig}
	\\
	\sigma_t 
	&= \frac{1}{\sqrt{\sum_i 1/\sigma_i^2}},
	\label{eq:weighted_integ}
\end{align}
where $I_i$ and $\sigma_i$ are the estimated value of the integral and standard deviation for the $i$-th iteration. The result for an iteration is weighted by $\sigma^{-2}$. Therefore, the iterations with smaller standard deviations are more important for the accumulated result.

\begin{table}[t]
\begin{center}
    \begin{tabular}{c|c}
    \hline
     Loss & Formula 
     \\\hline\hline \\ [-1em]
     rdkl & $ N^{-1}\sum_i^N \log(p_i/q_i)$
     \\[+0.5em] \hline \\ [-1em]
     fchi2 & 
     $N^{-1}\sum_i^N (q_i - p_i)^2/p_i^2$
     \\[+0.5em] \hline \\ [-1em]
     hel2 & 
     $N^{-1}\sum_i^N (\sqrt{q_i} - \sqrt{p_i})^2/p_i$ 
     \\[+0.5em] \hline \\ [-1em]
     jeff &
     $N^{-1}\sum_i^N (q_i - p_i)\log (q_i/p_i)/p_i$
     \\[+0.5em] \hline \\ [-1em]
     chra & 
     $ 
     4 \left(1- 
     N^{-1}\sum_i^N 
     q_i^{(1-\alpha)/2} 
     p_i^{(1+\alpha)/2} 
     / p_i\right)/(1-\alpha^2)
     $
     \\[+0.5em] \hline \\ [-1em]
     expn & 
     $
     N^{-1}\sum_i^N
     (q_i/p_i)\log^2(q_i/p_i)
     $
     \\[+0.5em] \hline \\ [-1em]
     var &
     $\begin{aligned}
         N^{-1} \sum_i^N \left( 
         \left( q_i/p_i \right)^2 - 
         N^{-1} \sum_i^N
         \left( {q_i}/{p_i} \right)^2 \right)
     \end{aligned}$
     \\[+0.5em] \hline \\ [-1em]
     abpr & 
     $
     \begin{aligned}
         &\frac{2}{(1-\alpha)(1-\beta)}
         N^{-1}\sum_i^N \frac{q_i}{p_i}
         \nonumber\\ 
         &\times
         (1-p_i/q_i)^{(1-\alpha)/2}
         (1-p_i/q_i)^{(1-\beta)/2}
     \end{aligned}
     $
     \\[+0.5em] \hline
\end{tabular}
\caption{Definitions of loss functions employed in this work. We set $\alpha = 0.5$ in ``chra'' and $\alpha=\beta=0$ in ``abpr''. Here $p_i$ is the probability of the $i$-th sample from nflow, while $q_i = |\psi(\vec{x}_i)|/\tilde{I}$ is proportional to the magnitude of the integrand $\psi(x_i)$. \label{losst}}
\end{center}
\end{table}

\begin{figure}[t]
	\centering
	\includegraphics[width=\linewidth]{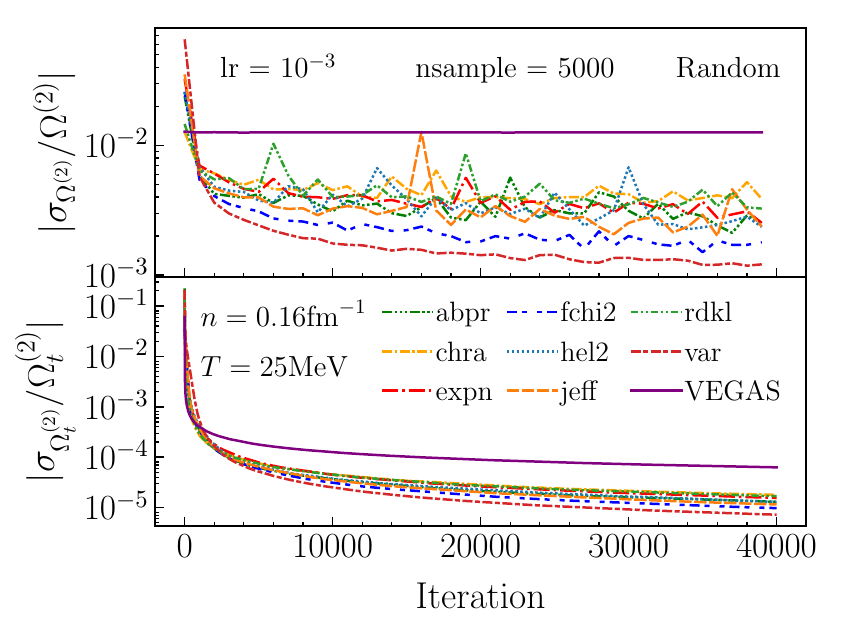}
	\caption{(Top panel) Batch relative error averaged over 1000 training iterations for the calculation of the second-order perturbative contribution to the grand canonical potential employing eight different choices of the loss function. (Bottom panel) Total relative error as a function of training iteration. In both cases, each iteration batch consists of 5000 samples. For comparison, we also show the results from VEGAS. In all cases, during training we employ the pseudo-random number sampler.}
	\label{fig:difloss}
\end{figure}

In Figure \ref{fig:difloss}, we show the relative error (top panel) averaged over 1000 iterations as well as the accumulated estimate of the integral (bottom panel) up to the last iteration shown. In each iteration, 5000 samples are drawn from a uniform distribution and passed through the normalizing flow transformation to produce the output sample points for the integral estimate in Eq.\ \eqref{mcie}. We also show for comparison the results obtained from the Monte Carlo importance sampling algorithm VEGAS. Early in the training, VEGAS is able to obtain a better integral estimate, but by the 1000th iteration normalizing flows have obtained a more accurate sampling distribution, regardless of the choice of loss function. For this seven-dimensional integral, we also observe that VEGAS converges to its precision limit within about 100 iterations. Normalizing flows, on the other hand, keep improving the model of the target distribution over many iterations. In the optimal cases of the ``fchi2'' and ``var'' loss functions, we see that the normalizing flow model outperforms the VEGAS importance sampling method by approximately an order of magnitude in the batch uncertainty by the end of training. Early in the training, we see that the choice of the loss function does not have a strong impact on the efficiency of the normalizing flow model. However, at large iterations, the performance of the ``fchi2'' and ``var'' loss functions are notably stronger than the others. Due to the factor of $p_i^2$ in the denominators, both loss functions tend to enhance the contributions where the distribution is small. The end result is that the loss functions can more quickly obtain a detailed description of the target integrand. An additional benefit of the ``var'' loss function is that the fluctuations in the uncertainty are significantly smaller than those from all other loss functions.

We next consider different methods for sampling random numbers within the integration domain. Since the base distribution for the normalizing flow is just a uniform distribution over the domain space, one can employ either pseudo-random number algorithms or quasi-random number sequences, the latter having the potential advantage that they more evenly cover the domain space and therefore may be more suitable for training the machine learning model. In particular, as the temperature is lowered, the Fermi-Dirac distribution functions in the calculation of the grand canonical potential become steep, and in this case quasi-random number sequences may have advantages over pseudo-random number sampling. In this paper, we consider Sobol, Halton, and Lattice as the algorithms for the quasi-random sample generators. The pseudo-random and Sobol sample generators are from PyTorch, while the Halton and Lattice sample generators are from the qmcpy library  \cite{choi_quasi-monte_2021}.

\begin{figure}[t]
	\centering
	\includegraphics[width=\linewidth]{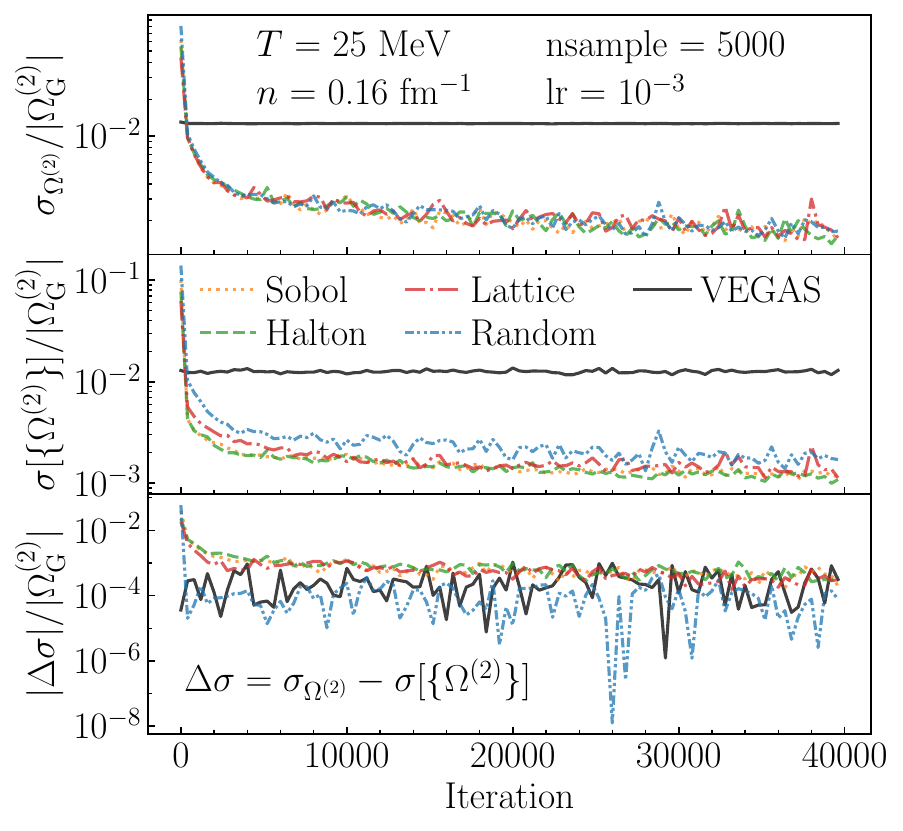}
	\caption{Different random number samplers are employed for training nflow. The lines ``Halton'', ``Sobol'', and ``Lattice'' correspond to different quasi-random sample generators coupled with the normalizing flow model, while ``Random'' is a pseudo-random sample generator. $\sigma_{\Omega^{(2)}}$ refers to the average standard deviation calculated using Eq.~\ref{eq:sig}. $\sigma[\{\Omega^{(2)}\}]$ refers to the standard deviation of the evaluated results over every 100 iterations. In all cases, the ``fchi2'' loss function is employed.}
	\label{fig:train_E2_sample}
\end{figure}

\begin{figure}[t]
    \centering
    \includegraphics[width=1.0\linewidth]{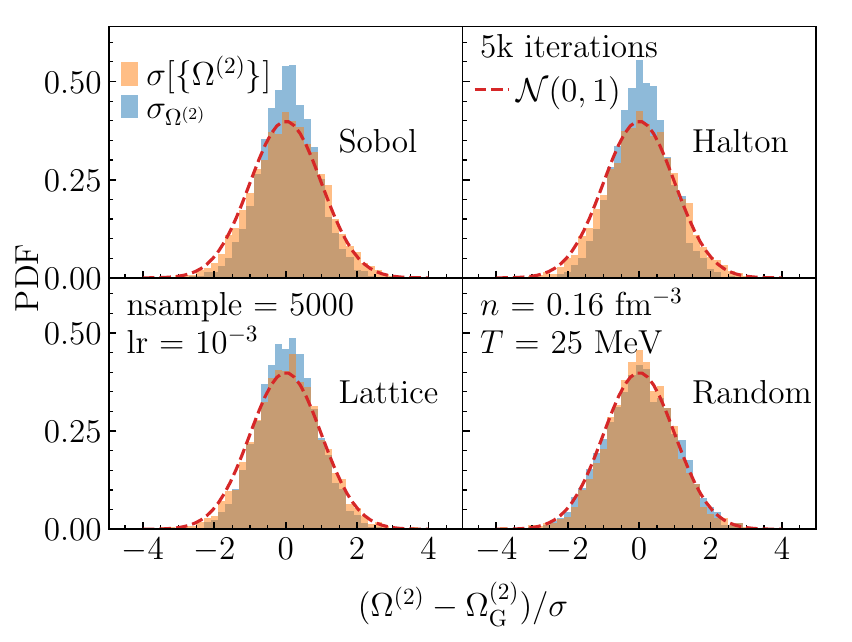}
    \caption{Normalized distribution of the evaluated $\Omega^{(2)}$ over the last 5000 training iterations with different sample generators. $\sigma_{\Omega^{(2)}}$ (blue) is the evaluated standard deviation using Eq.~\ref{eq:sig} based on samples at each iteration. $\sigma[\{\Omega^{(2)}\}]$ (orange) is the standard deviation of the set of evaluation results for all the last 5000 training iterations. The standard normal distribution (red dashed lines) is plotted for comparison.}
    \label{fig:gausstest}
\end{figure}

In the top panel of Figure \ref{fig:train_E2_sample} we show the batch relative uncertainty averaged over 1000 training iterations for the grand canonical potential of symmetric nuclear matter at density $n=0.16\,{\rm fm}^{-3}$ and temperature $T=25\,{\rm MeV}$ calculated from normalizing flow models based on pseudo-random number and quasi-random number generators. At first sight, it appears that there is almost no dependence of the convergence on the choice of random number generator. However, in the middle panel of Figure \ref{fig:train_E2_sample} we show the same relative error as a function of iteration, except that the error is defined as the standard deviation of integral estimates over 1000 iterations (rather than the point estimate of Eq.\ \eqref{eq:sig} that can be obtained at each iteration). Instead, we now see that all of the quasi-random number generators offer a statistically significant improvement in relative error. Finally, in the bottom panel of \ref{fig:train_E2_sample} we show the difference in the two definitions of the relative error estimate as a function of iteration. One sees that for ``Random'' and ``VEGAS'', which are both obtained using a pseudo-random number generator, the two definitions of the uncertainty are equal to within 10\%. However, for the quasi-random number generators, the point estimate of the relative uncertainty $\sigma_{\Omega^{(2)}}$ is significantly larger, by up to 100\%, than the standard deviation of the integral estimates. Therefore, it appears that the point estimate uncertainty might be misleadingly large.

To further investigate this difference, we show in Figure \ref{fig:gausstest} the distribution of the true uncertainty over 5000 iterations normalized by the point error estimate $\sigma_{\Omega^{(2)}}$ (blue) or the standard deviation of the integral estimate $\sigma \left[ \{ \Omega^{(2)} \} \right ]$ (orange) for each of the random number generators. For reference, we also show as the dashed red lines a normal distribution with mean 0 and standard deviation 1. To obtain the true values of the integrals, we have employed Gaussian quadrature that is well converged as the number of integration mesh points is increased. From Figure \ref{fig:gausstest}, one finds that there is essentially no difference in these two distributions in the case of the pseudo-random number generator. However, for each of the quasi-random number generators, the true distribution of errors does not match the point estimates since the blue curves deviate strongly from the red dashed normal distributions. On the other hand, the orange distributions that are obtained from the variation in integral estimates follow a normal distribution. We conclude from Figures \ref{fig:train_E2_sample} and \ref{fig:gausstest} that (i) the quasi-random number generators provide more precise integral estimates than their point distributions would suggest and (2) the quasi-random number generators lead to faster convergence and more precise integral estimates than pseudo-random number generators.

\begin{figure}
    \centering
    \includegraphics[width=\linewidth]{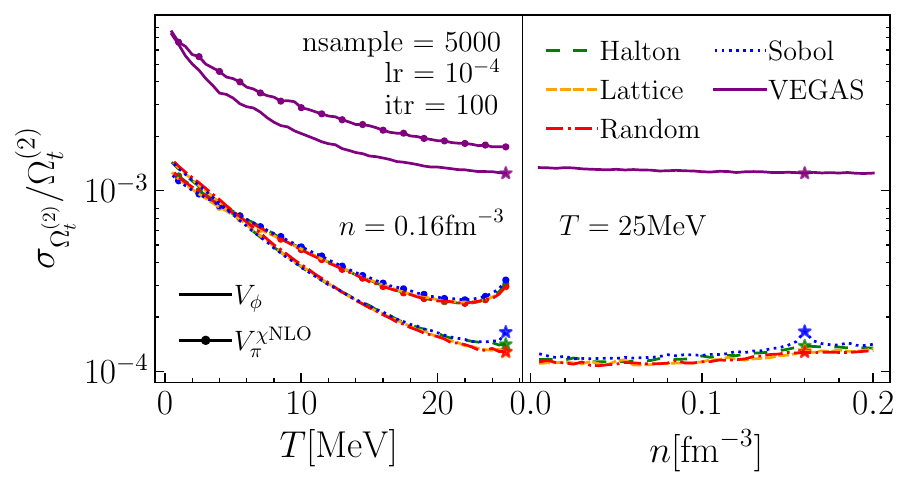}
    \caption{Batch relative error for $\Omega^{(2)}$ as the temperature (left) and density (right) are varied. The lines with solid circles denote the replacement of the one-boson exchange interaction of Eq.\ \eqref{obe} with the non-trivial $\pi$-exchange interactions up to NLO from chiral effective field theory. In all cases, we start from the highly-trained models from Figure \ref{fig:train_E2_sample}, shown here with a star.}
    \label{fig:transfer}
\end{figure}

After training a normalizing flow model with a particular nuclear force model at a specific density and temperature, we investigate the efficiency of the pre-trained model when transferred to a nearby phase space point and using a different nuclear potential. After each new modification, we perform a 100-iteration update for the new models. Then the weighted estimation for the 100 iterations is calculated. We show the results in Figure \ref{fig:transfer}, where we plot the batch relative uncertainty using 5000 sample points as the temperature (left panel) and density (right panel) are varied. The star denotes the starting value of the pre-trained model. For comparison we also show the results from VEGAS as the purple lines. First, we notice that as the temperature is lowered, the uncertainties in the calculation of the grand canonical potential increase for both normalizing flows and VEGAS. This is to be expected since the Fermi-Dirac distribution functions gradually approach step functions with steeper integrands within the integration region. In contrast, as can be seen in the right panel, for a fixed temperature and varying density the relative uncertainty remains roughly constant. In all cases, normalizing flows retain an approximate one order of magnitude improvement over VEGAS in the relative uncertainty.

In Figure \ref{fig:transfer}, we also show with the dotted lines the performance of the pre-trained model to estimate $\Omega^{(2)}$ when the simple scalar-isoscalar boson-exchange model is substituted with a more realistic next-to-leading-order pion-exchange model ($V_\pi^{\chi \rm NLO}$) from chiral effective field theory. We still find a one-order magnitude of improvement for normalizing flows when compared with VEGAS for the transfer calculation. However, both Monte Carlo importance sampling methods exhibit an increase in the relative error estimate when the chiral pion-exchange potential is employed.

\begin{figure}[t]
    \centering
    \includegraphics[width=0.9\linewidth]{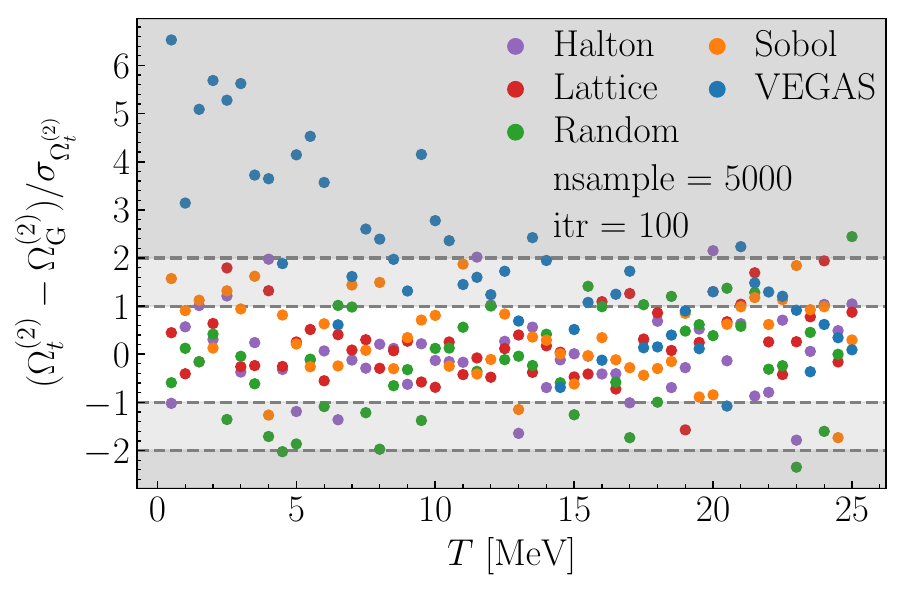}
    \caption{Ratio of the actual error to the variance-estimated error from different random number generators. The exact result is obtained from very precise Gaussian quadrature integration.}
    \label{fig:difTcompare}
\end{figure}

To end this subsection, we show in Figure \ref{fig:difTcompare} the ratio of the true error in the Monte Carlo integral estimate of $\Omega^{(2)}$ to the predicted error from the importance sampling variance $\sigma_{\Omega^{(2)}}$. From Figure \ref{fig:difTcompare} we see that for temperatures above $T=10$\,MeV, all importance sampling algorithms produce accurate integral estimates that fall within one or two standard deviations of their predicted errors. However, at lower temperatures one finds that normalizing flows continue to accurately estimate their errors while VEGAS tends to underpredict its error. In this regime, the VEGAS integral estimates routinely lie beyond three standard deviations from the true results.

\subsection{Density-density response function}
\begin{figure}
	\centering
	\includegraphics[width=0.3\linewidth]{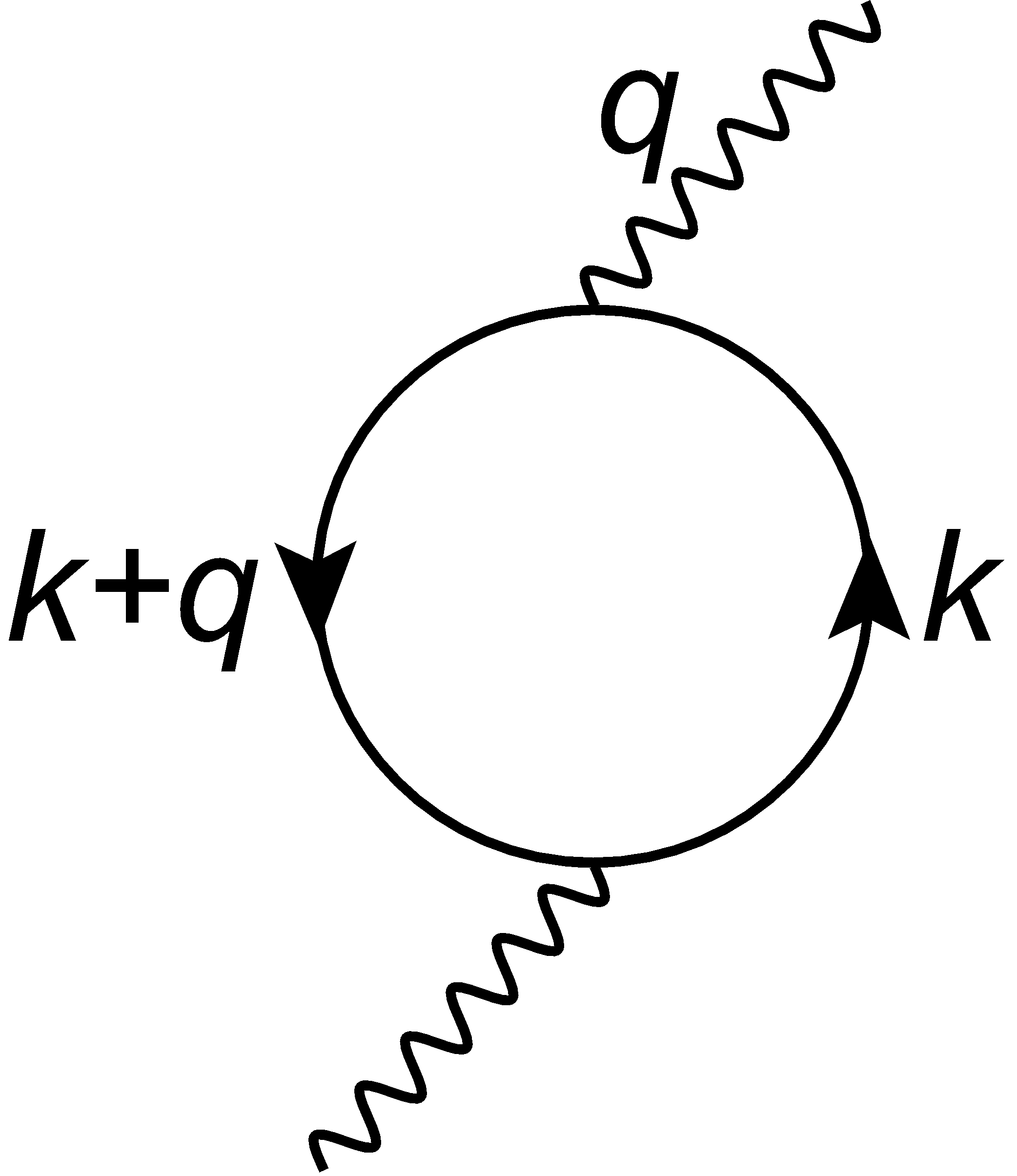}
	\caption{Zeroth-order static density-density response function. 
	The solid and wavy lines represent baryon and external perturbation separately, where $q$ is the external perturbative momentum. }
	\label{fig:0th-order_response_function}
\end{figure}

In this subsection, we investigate the ability of normalizing flows to model integrands that appear in the calculation of complex-valued nuclear matter response functions and especially the principal value integrals needed to calculate the real parts. Specifically, we consider the density-density response function of symmetric nuclear matter at zero temperature. The density-density response function measures how the density of nuclear matter is influenced by an external perturbation and is defined as
\begin{align}
	\chi (t_2, t_1)
	\equiv -i 
	\langle T 
	\hat{\psi}^\dagger(t_1) \hat{\psi}(t_1)
	\hat{\psi}^\dagger(t_2) \hat{\psi}(t_2)
	\rangle.
\end{align}
In the momentum and frequency representation, the zeroth-order density-density response function can be represented by the Feynman diagram in Figure \ref{fig:0th-order_response_function}. For symmetric nuclear matter, it is given by
\begin{align}
	\chi (q, \omega)
	&= 
	4 
	\int \frac{d^3 k}{(2\pi)^3}
	\Bigg[
		\frac{(1 - f_{\pvec{k}+\pvec{q}} ) f_{\pvec{k}}}
		{\omega + \epsilon_{\pvec{k}} - \epsilon_{\pvec{k} + \pvec{q}} + i\eta}
		\nonumber\\
		&\quad
		- 
		\frac{(1 - f_{\pvec{k}} ) f_{\pvec{k}+\pvec{q}}}
		{\omega + \epsilon_{\pvec{k}} - \epsilon_{\pvec{k} + \pvec{p}} - i\eta}
	\Bigg],
    \label{eq:response}
\end{align}
where $\pvec{k}$ is the momentum of the incoming particle, $\pvec{q}$ is the momentum transferred to the particle by the external probe, and $\omega$ is the energy transfer. The Fermi-Dirac distribution function is denoted by $f_{\pvec{k}} = \Theta( k_F - k)$ at zero temperature, where $k_F$ is the Fermi momentum. 

The response function can be decomposed into real and imaginary parts using the relation 
$\frac{1}{\omega\pm i\eta} = \mathcal{P}\frac{1}{\omega}\mp i\pi\delta(\omega)$, where ${\cal P}$ denotes the principal value integral. In the kinematic region where there is an imaginary part, the real part is then given by
\begin{align}
	&\quad 
	\Re \chi\left(q>2 k_F, 
		\omega \in \left( \frac{q^2 + qk_F}{2m}, \frac{q^2 + 2qk_F}{2m} \right) 
		\right)
	\nonumber\\
	&= 
	- \frac{2m}{\pi^2}
	\mathcal{P}
	\int_{-1}^{1} d \! \cos\theta \int_0^{k_F} dk 
	F_{\chi}(k, \theta, q, \omega, m),
	\label{eq:Fchi}
\end{align}
where the integrand is 
\begin{align}
	&F_{\chi}(k, \theta, q, \omega, m)
	= 
	\frac{k^2}{q^2 + 2kq\cos\theta - 2m\omega}
	\nonumber\\
	&\qquad + 
	\frac{k^2}{q^2 + 2kq\cos\theta + 2m\omega}.
\end{align}
There is a pole line $k_{\rm pole}$ in the integration region that leads to the divergence of the integrand along the line:
\begin{align}
	k_{\text{pole}} = \frac{2m\omega - q^2}{2q\cos\theta}.
\end{align}
Two strategies are widely applied to deal with poles for Cauchy principal value of integrals: Reflection and Subtraction. 
We implement these two strategies with Monte Carlo integration and compare their performance. We will find that normalizing flows have the ability to be applied in both cases.

\begin{figure}
	\centering
	\includegraphics[width=0.9\linewidth]{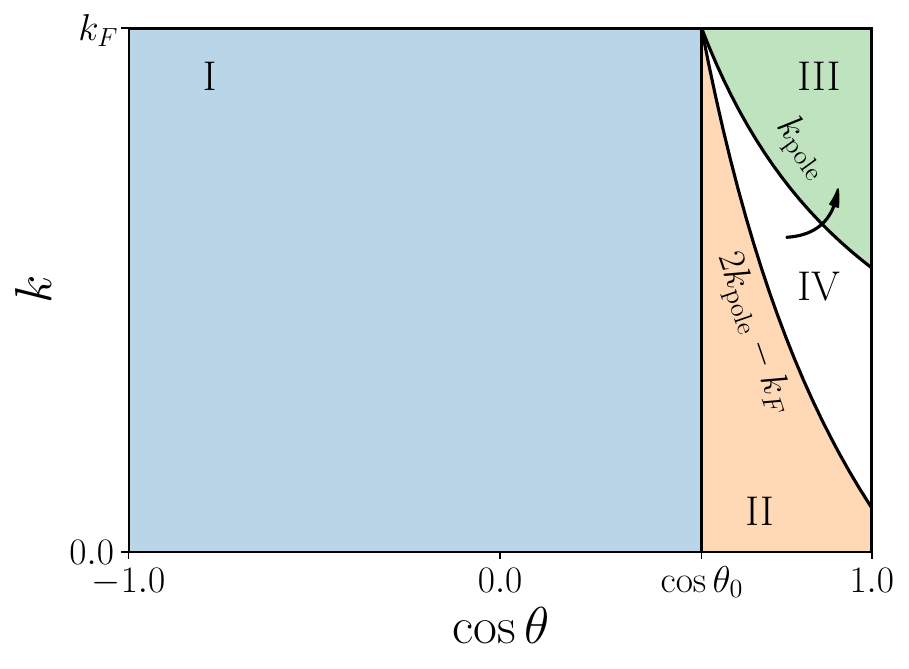}
	\caption{Integration regions for the $0^{\rm th}$-order density-density response functioin. $k_{\rm pole}$ is pole line where the integrand divergent. The curved arrow indicates the integrand in region \uppercase\expandafter{\romannumeral4} is flipped up into the region \uppercase\expandafter{\romannumeral3}.}
	\label{fig:ResFIntegrationRegion}
\end{figure}

\paragraph{Reflection}
The reflection strategy flips the integrand from one side near of pole into another side by changing the coordinate. 
The flipped integrand then is combined with the unflipped integrand as
\begin{align}
	&\quad 
	\int^{k_{\rm pole}} dk F(k) + \int_{k_{\rm pole}} dk F(k)
	\nonumber\\
	&= 
	\int_{k_{\rm pole}} dk F(2k_{\rm pole} - k) + F(k).
\end{align}
The positive and negative divergences are canceled after the recombination.
\begin{align}
	&\quad
	\lim_{k \to k_{\rm pole}}
	F(2k_{\rm pole} - k) + F(k)
	\nonumber\\
	&=
	\frac{20 m^2 \omega^2 - 12 m \omega q^2 + q^4}{8m \cos^2\theta q^2}.
\end{align}
As shown in Figure \ref{fig:ResFIntegrationRegion}, the integrand in region \uppercase\expandafter{\romannumeral4} is flipped up and recombined with region \uppercase\expandafter{\romannumeral3} so that the recombined integrand is finite along the line $k_{\rm pole}$. After the reflection, there are three independent integrals corresponding to the integration regions \uppercase\expandafter{\romannumeral1}, \uppercase\expandafter{\romannumeral2}, and \uppercase\expandafter{\romannumeral3}. Therefore, we can apply three independent normalizing flows or VEGAS \cite{lepage_adaptive_2021} to calculate them and sum over them at the end to get the value of the integral. 

\begin{figure*}[t]
	\centering
	\includegraphics[width=\linewidth]{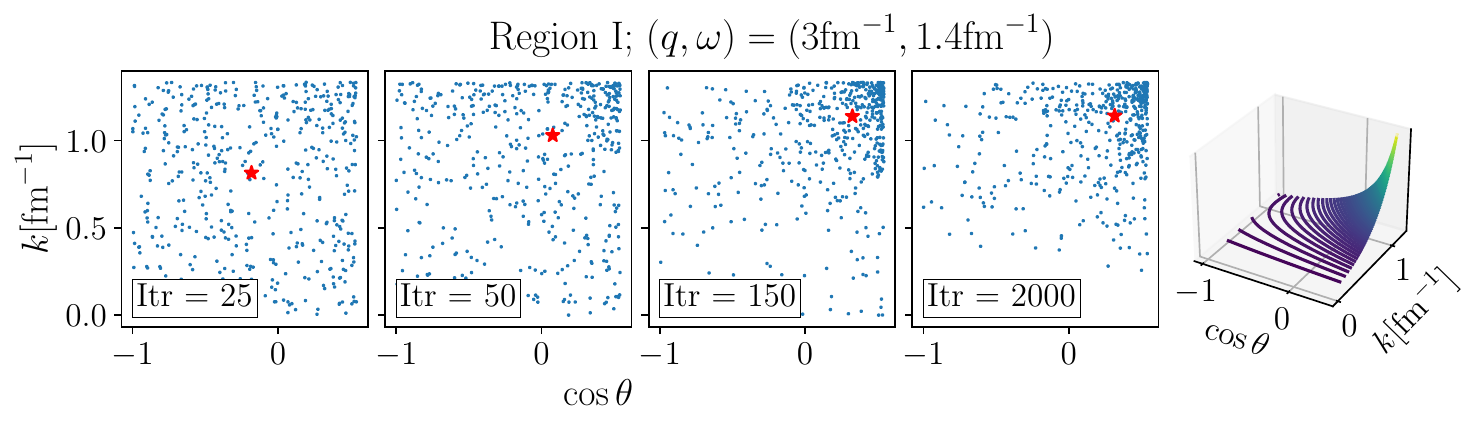}
	\caption{The distribution of transformed samples at different training iterations within Region \uppercase\expandafter{\romannumeral1}. 
	The right 3D figure shows the integrand in Eq.~\ref{eq:Fchi}. 
	The initial learning rate for this training is $1\times 10^{-3}$.
	The star is the transformed sample generated from the fixed point at the center of the integration region.}
	\label{fig:SampleDistribution}
\end{figure*}

\paragraph{Subtraction:}
The subtraction strategy combines two divergent integrands in such a way that the sum is finite across the pole line. This method only needs to consider the single integration area composed of Regions \uppercase\expandafter{\romannumeral2}, \uppercase\expandafter{\romannumeral3}, and \uppercase\expandafter{\romannumeral4} shown in Figure \ref{fig:ResFIntegrationRegion} at the same time. In the case of the density response function, one has
\begin{align}
	&\quad 
	\int_{\cos\theta_0}^{1} d\cos\theta
	\int_{0}^{k_F} dk 
	\left[F(k) - \frac{k_{\rm pole}^2}{q^2 + 2kq\cos\theta - 2m\omega} \right]
	\nonumber\\
	&
	+ 
	\int_{\cos\theta_0}^{1} d\cos\theta
	\int_{0}^{k_F} dk 
	\frac{k_{\rm pole}^2}{q^2 + 2kq\cos\theta - 2m\omega}.
	\label{eq:subtraction}
\end{align}
Thus, the divergence vanishes for the first term in Eq.~\ref{eq:Fchi}, since $\lim_{k\to k_{\rm pole}}k^2 - k_{\rm pole}^2=0$ in the numerator. 
The second line in Eq.~\ref{eq:subtraction} can be calculated via $\int \frac{dx}{x - a} = \log |x-a|$ and leads to
\begin{align}
	&\quad
	\int_{\cos\theta_0}^{1} d\cos\theta
	\int_0^{k_F} dk 
	\frac{k_{\rm pole}^2}{q^2 + 2kq\cos\theta - 2m\omega}
	\nonumber\\
	&= 
	\int_{0}^1 dx
	\int_{\cos\theta_0} d\cos\theta
	k_{\rm pole}^2
	\nonumber\\
	&\qquad \times
	\log 
	\left(\left| \frac{q^2 + 2k_F q\cos\theta - 2m\omega}{2m\omega - q^2} \right|\right),
	\label{eq:logintegral}
\end{align}
where we have added an auxiliary one-dimensional integral $\int_0^1 dx$ here so that coupling transforms can be used as input to build the normalizing flow transformation functions. We therefore arrive at three independent integrals: the integration over Region \uppercase\expandafter{\romannumeral1}, the integration over Regions \uppercase\expandafter{\romannumeral2} - \uppercase\expandafter{\romannumeral4}, and the integration of the logarithm in Eq.\ \eqref{eq:logintegral}.

Both strategies require independent Monte Carlo integrators to calculate the corresponding integrals. The final result is the statistical summation of all the integrators:
\begin{align}
	I = \sum_{i=1}^3 I_i, \quad \sigma = \sqrt{\sum_{i=1}^3\sigma_i^2},
\end{align}
where $I_i$ and $\sigma_i$ are the statistical mean value and standard deviation from the Monte Carlo integrators for the $i$-th integral.

\subsection{Response functions: Sampling algorithms and loss functions}

In this subsection, we use normalizing flows to calculate the zeroth-order nuclear matter response function at zero temperature. We assess the performance of loss functions, random number sampling algorithms, and the ability of normalizing flow models to transfer to new values of the energy and momentum transfer. We apply both the Reflection and Subtraction strategies. As an illustration, in Figure \ref{fig:SampleDistribution} we show how the samples within Region \uppercase\expandafter{\romannumeral1} of Figure \ref{fig:ResFIntegrationRegion} are distributed at different times in the training. We choose symmetric nuclear matter at density $n=0.16\, {\rm fm}^{-3}$ and take $(q,\omega) = (3\,\text{fm}^{-1}, 1.4\,\text{fm}^{-1})$. We employ a pseudo-random number generator together with the ``fchi2'' loss function.
From Figure \ref{fig:SampleDistribution}, we see that at the beginning of the training, the transformation functions produce nearly uniform distributions. However, during training, the transformed samples gather at the important region where the magnitude of the integrand is large. 
At the 2000th iteration, the distribution of transformed samples is close to the normalized magnitude of the integrand as shown in the right 3D diagram.

\begin{figure}[t]
	\centering
	\includegraphics[width=1.02\linewidth]{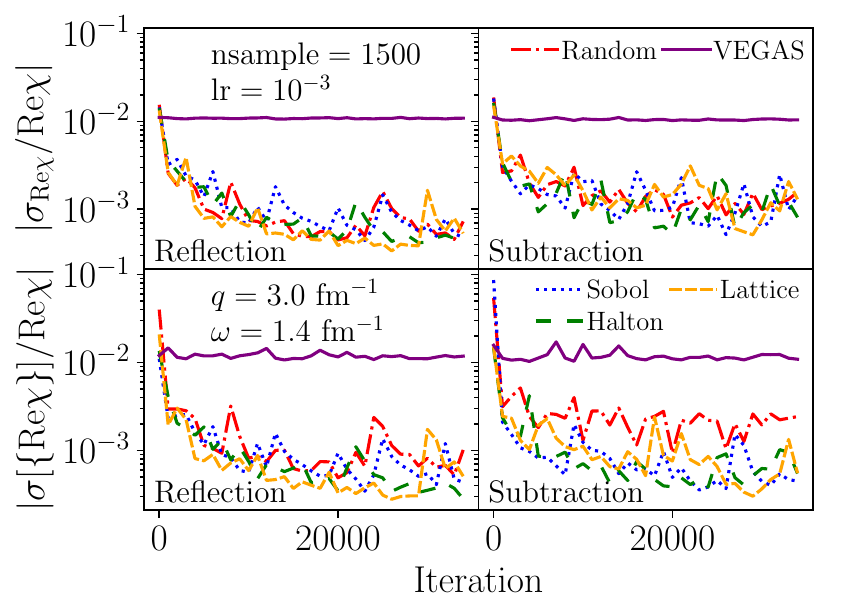}
	\caption{The average relative pointwise uncertainties (top) and standard deviation of integral estimates (bottom) over 1000 iterations for Monte Carlo importance sampling of the density response function of nuclear matter from normalizing flows and VEGAS with the Reflection (left) and Subtraction (right) algorithms. For normalizing flows, the initial samples in the uniform distribution are generated from quasi- and pseudo-random number algorithm. The quasi-random number algorithms are Sobol, Halton and Lattice.} 
	\label{fig:ResTaining}
\end{figure}

In Figure \ref{fig:ResTaining} we show in the top panels the pointwise relative error $\sigma_{{\rm Re}\, \chi} / {\rm Re}\, \chi$ in Eq.\ \eqref{eq:sig} for the density response function of symmetric nuclear matter at saturation density and for momentum transfer $q=3.0\,{\rm fm}^{-1}$ and energy transfer $\omega=1.4\,{\rm fm}^{-1}$ calculated from normalizing flow importance sampling with different choices for the sample generators: Sobol, Halton, Lattice, and pseudo-random. The uncertainties generated by normalizing flows are also compared to those from the VEGAS importance sampling algorithm. We consider both strategies, Reflection (left) and Subtraction (right), for evaluating the principal value integrals over regions that contain poles. We use a batch size of 1500 samples at each iteration and 35,000 iterations to train the Monte Carlo integrators. As shown in Figure \ref{fig:ResTaining}, VEGAS again converges quickly to its precision limit and has a better performance over normalizing flows at the beginning of the training. The relative standard deviation from VEGAS drops to $10^{-2}$ within the first 100 iterations but has no improvement during further training. However, the transformation functions generated by normalizing flows continue optimizing. Thus, the uncertainties from normalizing flows can become significantly smaller than those from VEGAS. At the end of the training, the relative uncertainty reached by the Reflection normalizing flow is less than $10^{-3}$, which indicates a more than one order of magnitude improvement over VEGAS in terms of the batch precision. We see that neither the overall pointwise precision nor the training speed depends significantly on the type of sample generators, as in the case of the grand canonical partition function.

In the lower panels of Figure \ref{fig:ResTaining}, we show the standard deviation of the integral estimates averaged over 1000 iterations. As in the case of the grand canonical potential, we find that the quasi-random number generators enable more stable training with a reduced relative uncertainty compared to the results using a pseudo-random number generator.
In both cases, we see that there is a moderate improvement in the overall efficiency when the Reflection algorithm is employed to treat the principal value integral.

\begin{figure}
	\centering
	\includegraphics[width=1.\linewidth]{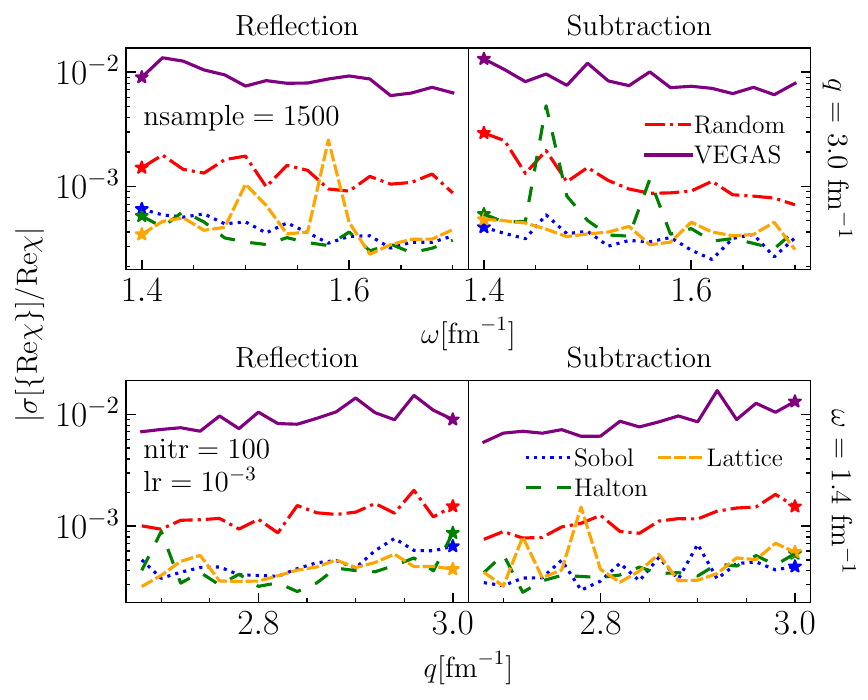}
	\caption{Standard deviation of Monte Carlo importance sampling integral estimates of the density response function of nuclear matter when a highly-trained model is transferred to new values of the energy transfer (top) and momentum transfer (bottom). All the normalizing flow and VEGAS models are trained at $({q}, \omega) = (3.0\,{\rm fm}^{-1}, 1.4\,{\rm fm}^{-1})$. The relative uncertainty is calculated via accumulated estimation for 100 iterations and 1500 samples per iteration. The points marked by stars indicate the values of the energy and momentum transfer for the highly-trained models.}
	\label{fig:Res_DifQOmega}
\end{figure}

After training the normalizing flow models at one position in the kinematic $(q,\omega)$ space, we can study the transferability of the highly-trained model to other values of the momentum and energy transfer. In particular, it is not clear a priori whether the pole structure and related principal value integrals will be efficiently transferred. We start by training the normalizing flow and VEGAS models at $({q},\omega) = (3.0\, {\rm fm}^{-1}, 1.4\, {\rm fm}^{-1})$ as shown in Figure \ref{fig:ResTaining}. We then transfer the well-trained models to calculate $\Re\chi$ at other $({q},\omega)$ positions as shown in Figure \ref{fig:Res_DifQOmega}.
At the beginning of the model transfer, the trained models are used to estimate the density response function for 100 iterations with 1500 samples per iteration at the $({q},\omega)$ position where they are trained. After the 100 iterations, new models are generated automatically and they are used to calculate the response function at the next position. This transfer process can be performed iteratively. 
As shown in Figure \ref{fig:Res_DifQOmega}, we find that the integrands at one value of the energy and momentum transfer are sufficiently similar to the integrands at nearby values to enable an efficient transfer of the highly-trained model to different regions of the kinematic space. During the transfer, both normalizing flows and VEGAS retain roughly the same precision as the original highly-trained model. The accumulated relative standard deviations at every position show an approximate order 
of magnitude improvement for normalizing flows over VEGAS. We see also that the quasi-random number generators similarly outperform the pseudo-random number generators as the energy and momentum transfer are varied.

\begin{figure*}[htp]
	\centering
	\includegraphics[width=0.75\linewidth]{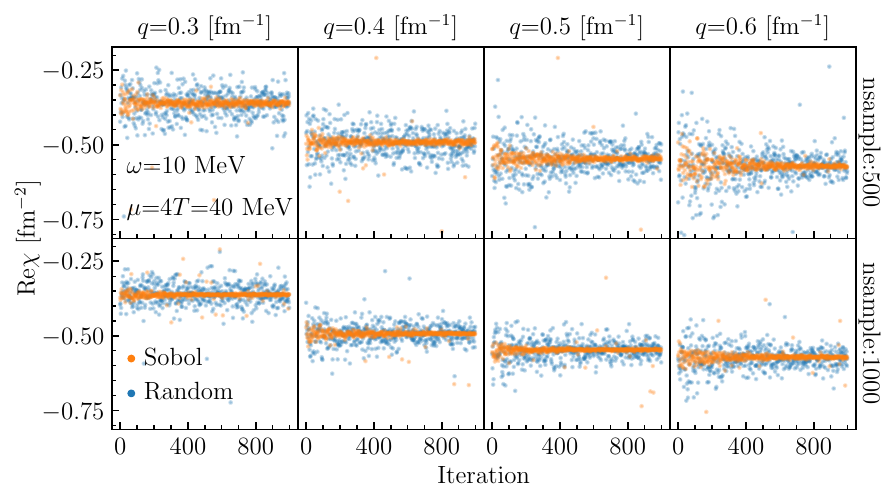}
	\caption{Monte Carlo estimate for the real part of 0th-order response function ${\rm Re}\chi$ at finite temperature with different external momenta (columns) and number of samples per batch (rows). 
    The samples are generated using pseudo-random number generators (blue dots) and quasi-random number generators (orange dots). 
	}
    \label{fig:sobol_response}
\end{figure*}


Finally, we investigate the efficiency of normalizing flow importance sampling for different sample generators when the number of samples is low. We anticipate that since the quasi-random number generators more evenly cover the domain space, they may have advantages over pseudo-random number generators when the number of samples is small. For this study, we consider the real part of the density response function of pure neutron matter at finite temperature. For the case $2m\omega > q^2$, there is a pole located in the region $\cos\theta>0$. In this case, the integral Eq.\ \eqref{eq:response} can be separated into three parts
\begin{align}
    &\Re\chi\left(q,\omega>\frac{q^2}{2m}\right) 
    = 
    \int_{-1}^0 d \cos\theta \int_0^\infty d k F_\chi^{(T)}(\cos\theta, k)
    \nonumber\\
    &\quad + 
    \int_{0}^1 d \cos\theta \int_0^{k_{\rm pole}} d k 
    \Big[F_\chi^{(T)}(\cos\theta, k) 
    \nonumber\\
    &\quad\qquad
    + F_\chi^{(T)}(\cos\theta, 2k_{\rm pole} - k)\Big]
    \nonumber\\
    &\quad + 
    \int_{0}^1 d \cos\theta \int_{2k_{\rm pole}}^\infty d k F_\chi^{(T)}(\cos\theta, k),
    \label{eq:responseT}
\end{align}
where the second part uses the reflection strategy to deal with the pole, and the integrand is given by 
\begin{align}
    F_\chi^{(T)}(\cos\theta, k)
    = \frac{m}{\pi^2 q}
    \frac{k^2(f_{\pvec{k}} - f_{\pvec{k}+\pvec{q}})}{\cos\theta(k_{\rm pole} - k)}.
\end{align}
The three integrals in Eq.\ \eqref{eq:responseT} can be recombined into one by changing the coordinates $(\cos\theta, q)$ into the unit square. In Figure \ref{fig:sobol_response} we show the values of ${\rm Re}\, \chi$ as a function of training iteration for fixed energy transfer $\omega = 10$\,MeV and different values of the momentum transfer $q = \{ 0.3, 0.4, 0.5, 0.6 \}$\,fm$^{-1}$ (columns) and number of samples ``nsample'' $= \{500, 1000\}$ (rows). The temperature is set at $T=10$\,MeV, and the chemical potential is fixed at $\mu = 40$\,MeV. The values of the response function obtained from the pseudo-random number generator are shown as blue dots, while the results from the Sobol quasi-random number generator are shown as orange dots. One first sees that for both values of ``nsample'', the Sobol quasi-random number generator outperforms the pseudo-random number generator. In addition, when the number of samples is reduced from 1000 to 500, the uncertainties from the pseudo-random number generator increase faster compared to the quasi-random number generator, as seen from the larger scattering of blue dots compared to orange dots. Therefore, the low-discrepancy property of the Sobol quasi-random number generator makes the integral estimation more stable through the training.

\section{Conclusion}
\label{sec:Conclusion}
In this work, we have illustrated the applicability of normalizing flows for Monte Carlo importance sampling of nuclear many-body perturbation theory calculations. The normalizing flow machine learning model is trained to learn a series of change-of-variable transformations that evolve a simple base distribution (e.g., a uniform distribution) to a complicated target distribution. We have shown that normalizing flows can indeed produce highly expressive probability density functions for importance sampling of integrands that appear in calculations of the nuclear matter equation of state and density response function. In particular, the density response function is a complex-valued quantity whose real parts must be evaluated with principal value integrals. We have found that the errors in equation of state and response function calculations are at least an order of magnitude smaller than those obtained from the VEGAS, a state-of-the-art importance sampling algorithm commonly used in the community. Although normalizing flows take significantly longer to train than VEGAS, we have shown that once a highly-trained normalizing flow model is obtained at one density/temperature point in phase space, it can be transferred to nearby phase space points without additional fine training. In addition, normalizing flows trained for calculations of the density response function could be used in other kinematic regions of energy and momentum transfer without significant loss of precision. Finally, a well-trained normalizing flow model can even accommodate variations in the choice of nuclear potential.

We have explored the use of different loss functions for training normalizing flow models and have also investigated the use of quasi-random number sequences for sample generation. We have found that quasi-random number generators are especially useful when small sample sizes are used in training. In addition, the true errors in many-body calculations using quasi-random number sequences are smaller than their point estimates, Eq.\ \eqref{eq:sig}, would suggest. In contrast, pseudo-random number generators have point estimate uncertainties that match well the true uncertainty. We therefore conclude that quasi-random number sequences are a more efficient tool for Monte Carlo importance sampling with normalizing flows compared to pseudo-random number generators.

Tabulations of the nuclear matter equation of state and response functions require the repeated evaluation of high-dimensional integrals over changing conditions of temperature, density, isospin asymmetry, and in the case of response functions also the energy and momentum transfer. Monte Carlo importance sampling with normalizing flows can therefore provide a significantly more efficient framework for microscopic calculations of nuclear physics inputs to astrophysical simulations of core-collapse supernovae, neutron star cooling, and neutron star mergers. Importance sampling with normalizing flows can also accommodate differing models of the nuclear force, which opens up the possibility for improved uncertainty quantification of nuclear matter quantities for astrophysical tabulations.

\bibliographystyle{apsrev4-1}

%

\end{document}